\documentclass[english]{report}
\usepackage[T1]{fontenc}
\usepackage[latin9]{inputenc}
\usepackage{geometry}
\usepackage{amssymb}
\usepackage{graphicx}
\usepackage{setspace}
\makeatletter

\providecommand{\tabularnewline}{\\}

\usepackage{cite}

\makeatother

\usepackage{babel}
\begin{document}

\title{Reliable Information Transmission along QCA Wires in the Presence
of Non-Adiabatic Transitions }

\author{Daniel Brox: EECE, University of British Columbia}
\maketitle
\begin{abstract}
\noindent Quantum dot cellular automata (QCA) computing schemes use
arrays of quantum dots as computational devices. Typically, these
operate ideally by maintaining arrays in their ground state to ensure
correct computational output. For large QCA circuits, thermal fluctuations
make this impossible, so adiabatic clocking has been proposed as a
means of dividing large circuit computations into subcircuit computations
that are more reliable. In this report, it is shown that wires and
inverters can transmit information correctly via their excited states
just as well as their ground states. A characteristic example of a
4 cell wire is simulated, and a theoretical derivation of this result
is given. When, non-nearest neighbor interactions are included in
the Hamiltonian, this result still holds true. On the other hand,
QCA majority gates and more complex circuits give incorrect results
when operated in excited states. These results suggest that gates
of reliable QCA circuits should be contained in smaller clocking zones,
while wires relaying information can be contained in larger clocking
zones. 
\end{abstract}

\section{Introduction}

In the drive to continue to increase computing power, methods of computing
based on quantum-dot cellular automata (QCA) have been proposed as
an alternative to conventional computing {[}1{]}. Rather than switching
between states of current flow, these methods work by manipulating
electronic configurations within atomic scale physical systems such
as groups of atoms or quantum dots, offering the potential of speeding
up and reducing the size of digital circuits {[}2{]}. Essential to
operation of QCA circuits is the switching of input cell states so
that new computations can be performed by the same circuit. This switching
must be performed with care, since it is desirable that QCA circuits
remain in their lowest energy state throughout operation to ensure
the correct electronic states of output cells. As a means of achieving
this, adiabatic clocking schemes have been proposed whereby tunneling
between quantum states of all non-input QCA cells is enabled while
the polarization of the input cells is switched, and tunneling is
disabled when switching is completed. In so doing, the adiabatic theorem
of quantum mechanics guarantees that if an energy gap is maintained
between the ground and excited states of the entire QCA circuit and
switching is slow enough, the circuit will stay in its ground state
with high probability through switching and the correct computation
will be performed {[}3{]}. 

However, there are still theoretical questions to be answered regarding
how well we should expect this technique to work for a particular
physical implementation when the environment of a circuit affects
its operation. For instance, a silicon dangling bond QCA circuit will
consist of cells fabricated on the surface of a silicon wafer interacting
with thermal vibrations in the silicon lattice that can potentially
flip the states of of its cell states via phonon emission / absorption
{[}4{]}. This is indicated in Fig. 1, where the third cell of a four
cell wire has switched its electronic configuration by absorbing a
phonon.

\begin{center}
\includegraphics[scale=0.4]{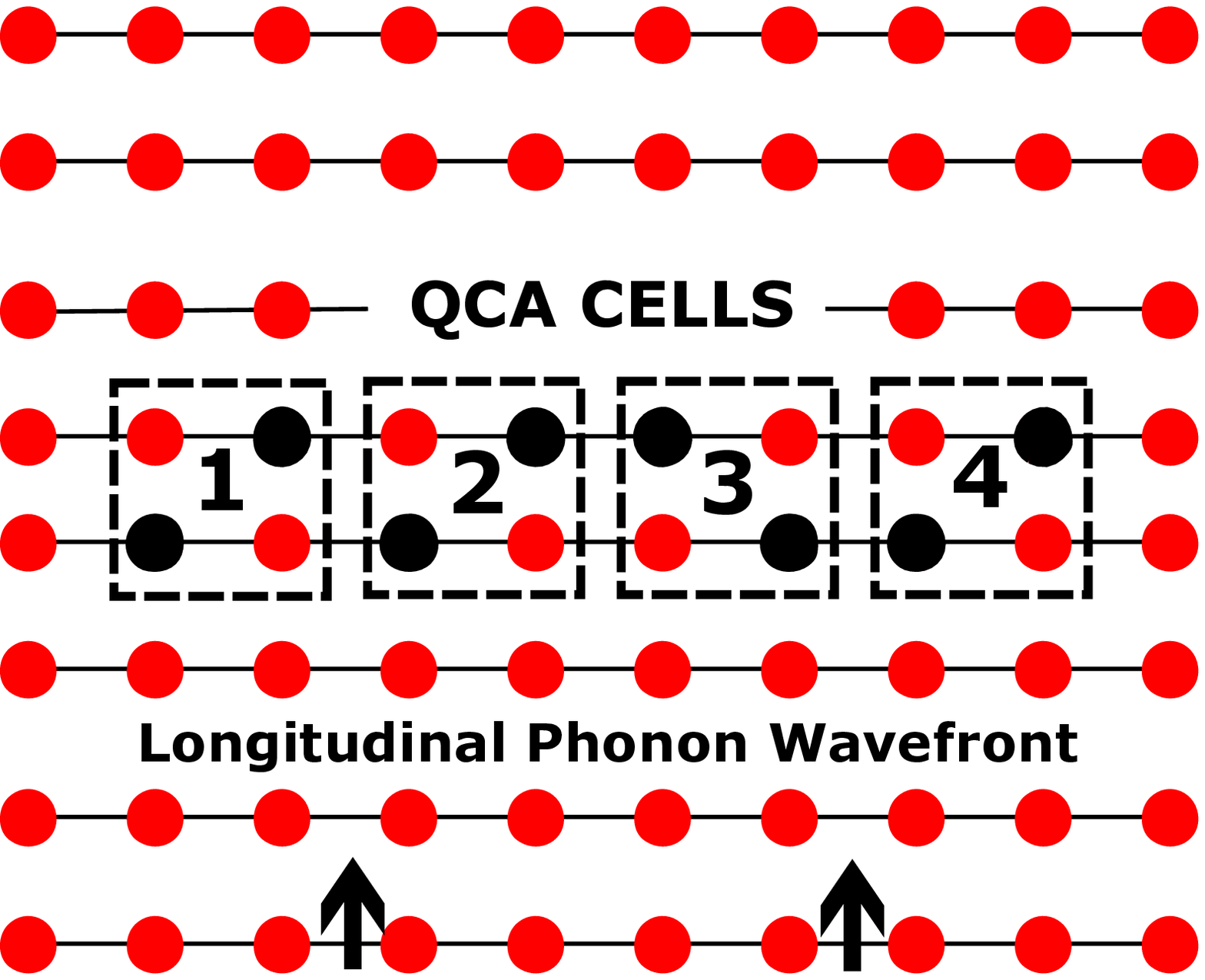}
\par\end{center}

\begin{center}
{\small{}Fig. 1 State flipping of a silicon dangling bond qubit due
to interaction with longitudinal phonons in a silicon lattice. Localized
electrons are shown as black dots and silicon atoms as red dots. In
this image, qubit 3 has flipped its electronic state by absorbing
a phonon.}
\par\end{center}{\small \par}

In general, excitation of a QCA circuit out of its ground state by
its environment can cause computation to give an incorrect answer.
This event becomes more and more likely as the number of cells in
the QCA circuit increases, so it has been suggested that dividing
a QCA circuit into separately clocked zones with small numbers of
cells will allow each zone to perform with high fidelity independently
of the others. If this is true, the probability of error occuring
in any particular zone circuit during its computation cycle can be
made small by sufficiently reducing the number of cells in the zone
and ensuring the energy gap between the ground state and first excited
state of each clocking zone circuit is large compared to the thermal
energy scale $k_{B}T$. Of course, in a sufficiently large QCA circuit,
a fraction of the zone circuits will not be in their ground states,
and we cannot assume that they will compute correctly as a consequence
of the adiabatic theorem. However, from a computational perspective,
we only care about whether or not a circuit gives the correct output
for a given set of computation inputs, and not about what quantum
state it starts out in. Therefore, the question arises as to whether
or not excited states of zone circuits can be utilized for computation,
as if this is possible, the reliability of the circuit is enhanced
and the amount of error correction in the overall circuit may be reduced
{[}5{]}. Initially, one might expect that the answer to this question
is no, since non-adiabatic transitions between excited states of an
adiabatically clocked circuit may randomize its output. However, for
the case of a QCA wire, the structure of the Hamiltonian describing
its quantum evolution enables information to be transmitted reliably
via excited states {[}6{]}. This finding has implications for QCA
circuit design, since it suggests that a lower density of clocking
zones may be used to relay information between different regions of
a circuit where a higher density of gate containing zones should be
used to perform computation.

\section{QCA Wire}

A simple simulation of a 4 cell QCA wire deomonstrates the phenomenon
of interest. Fig. 2 shows two possible wire starting states from which
we'd like to switch the input of the wire adiabatically to transmit
information:

\begin{center}
\includegraphics[scale=0.24]{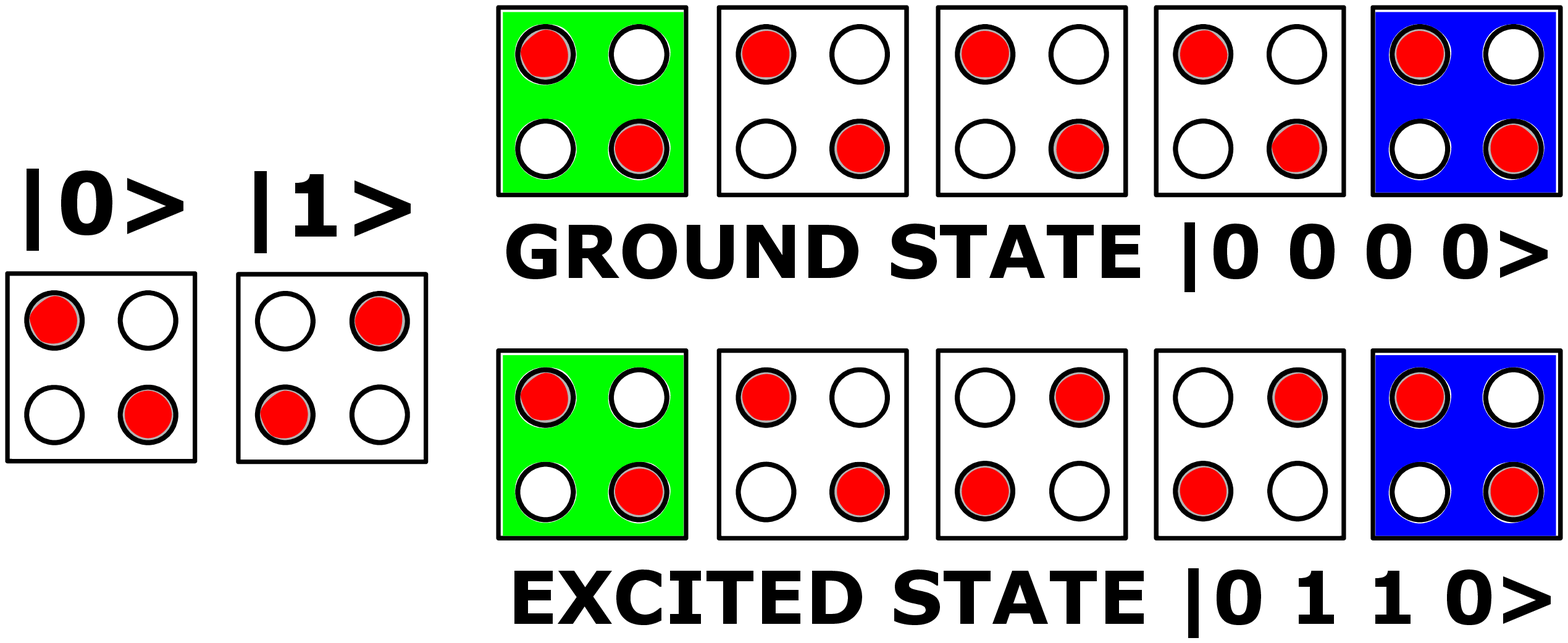}
\par\end{center}

\begin{center}
{\small{}Fig. 2 Schematic of ground and excited states of a four cell
QCA wire.}
\par\end{center}{\small \par}

\noindent In these wire states, each cell exists in a definite $|0\rangle$
or $|1\rangle$ state and the overall wire state is the tensor product
of 4 such states. The green input cell state is treated as a continuously
varying polarization $P_{in}(t)$. This is indicated in the QCA wire
Hamiltonian: 

\begin{eqnarray}
H_{wire}(t) & = & -\frac{E_{k}}{2}P_{in}(t)\sigma_{z}^{1}+\frac{E_{k}}{2}\gamma(t)(\sigma_{x}^{1}+\sigma_{x}^{2}+\sigma_{x}^{3}+\sigma_{x}^{4})\nonumber \\
 &  & -\frac{E_{k}}{2}(\sigma_{z}^{1}\sigma_{z}^{2}+\sigma_{z}^{2}\sigma_{z}^{3}+\sigma_{z}^{3}\sigma_{z}^{4}),
\end{eqnarray}

\noindent where $E_{k}$ is the kink energy associated with neighboring
cells having different polarizations (ie. $|0\rangle$ and $|1\rangle$
states), and $\gamma(t)$ is the adiabatic clocking signal that enables
/ disables cell tuneling. More specifically, when $|\gamma(t)|\approx1$,
cell tunneling is enabled and the cell state is ``unlatched'', and
when $|\gamma(t)|\approx0$, cell tunneling is disabled and the cell
state is ``latched''. Note that because there is only one adiabatic
clocking signal $\gamma(t)$ in the Hamiltonian, we are modeling the
wire as a singly clocked zone. 

The operators $\sigma_{z}^{i}$ and $\sigma_{x}^{i}$ in the Hamiltonian
are Pauli matrices {[}7{]} acting on the $i^{th}$ cell, which is
mathematically equivalent to a qubit in the two state approximation.
The $\sigma_{z}^{i}$ operators in the Hamiltonian are referred to
as polarization operators, and defined so that the cell states $|0\rangle$
and $|1\rangle$ are eigenvectors with eigenvalues -1 and 1. For example:

\begin{eqnarray}
\sigma_{z}^{1}|0\,1\thinspace1\thinspace0\rangle & = & -|0\,1\thinspace1\thinspace0\rangle,\\
\sigma_{z}^{2}|0\,1\thinspace1\thinspace0\rangle & = & |0\,1\thinspace1\thinspace0\rangle,\\
\sigma_{z}^{3}|0\,1\thinspace1\thinspace0\rangle & = & |0\,1\thinspace1\thinspace0\rangle,\\
\sigma_{z}^{4}|0\,1\thinspace1\thinspace0\rangle & = & -|0\,1\thinspace1\thinspace0\rangle.
\end{eqnarray}

\noindent Simulation of adiabatic switching is achieved by solving
the time dependent Schrodinger equation:

\begin{equation}
i\hbar\frac{\partial}{\partial t}|\Psi(t)\rangle=H_{wire}(t)|\Psi(t)\rangle,
\end{equation}

\noindent from $t=0$ to some final time $t=t_{f}$ with some choice
of functions $P_{in}(t)$ and $\gamma(t)$ such that $P_{in}(t_{f})=-1$,
$\gamma(t_{f})=0$, and the gap energy between the wire ground and
first excited state is nonzero throughout the switching interval.
For this purpose, we can set:

\begin{eqnarray}
P_{in}(t) & = & \cos(\pi t/t_{f}),\\
\gamma(t) & = & \gamma_{max}\cdot\sin(\pi t/t_{f}).
\end{eqnarray}

\noindent We then have two remaining parameters $t_{f}$ and $\gamma_{max}$
to choose. Since $\gamma_{max}$ will be chosen to maintain the gap
energy between the ground and first excited state close to the kink
energy $E_{k}$, it follows that we should choose $t_{f}\gg\hbar/E_{k}$
to avoid non-adiabatic transitions from the ground state to the first
excited state. Selecting $t_{f}=30\hbar/E_{k}$ to satisfy this condition,
we can plot the instantaneous energy spectrum (first 4 levels) of
the QCA wire for each time $t\in[0,t_{f}]$ for different values of
$\gamma_{max}$ to determine a suitable value. Fig. 3 shows three
such plots for $\gamma_{max}$= 0.1, 0.5, and 2.0, with energy in
units of $E_{k}$ and time in units of $\hbar/E_{k}$.

\begin{center}
\includegraphics[scale=0.64]{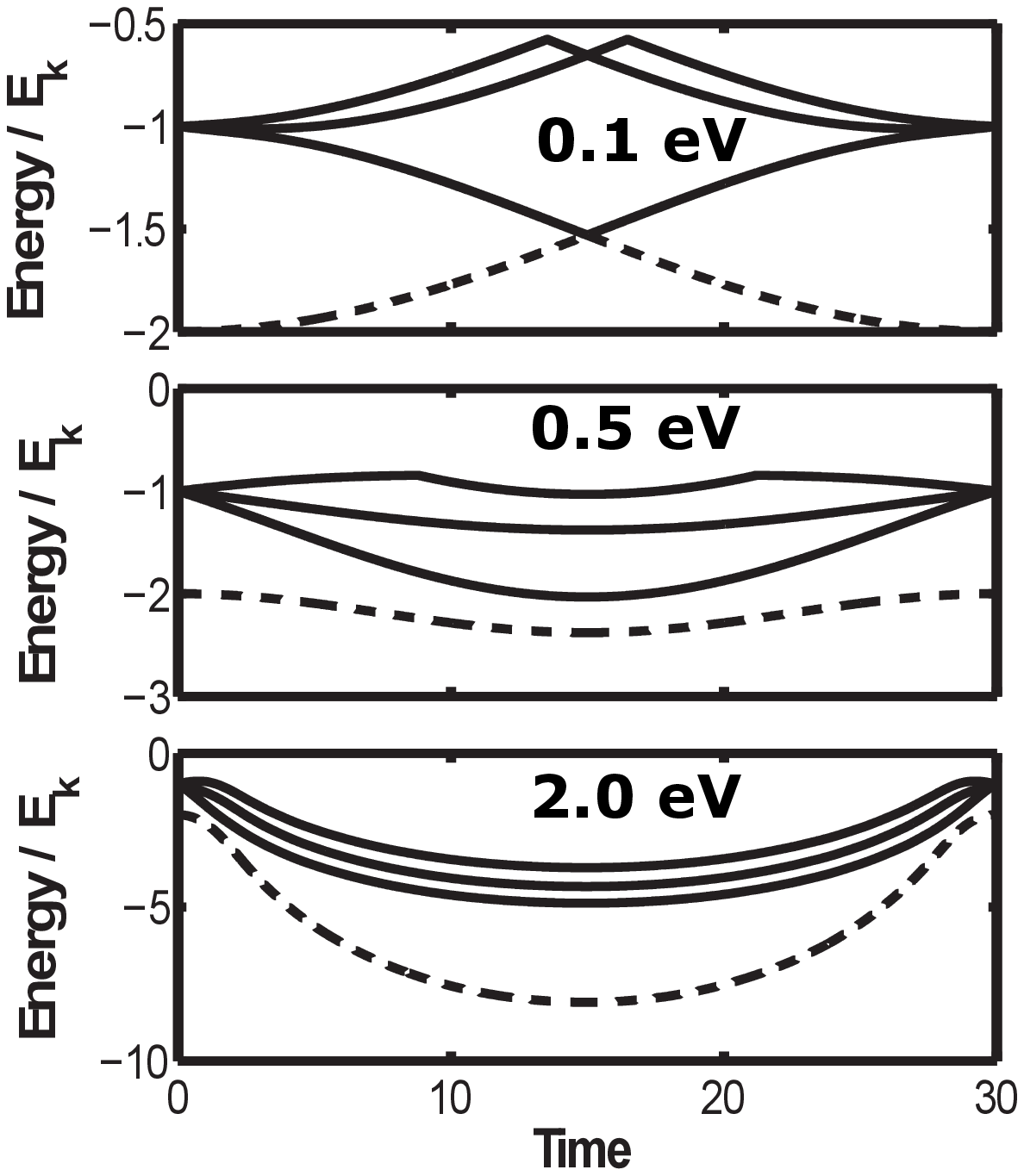}
\par\end{center}

\begin{center}
{\small{}Fig. 3 Four cell QCA wire energy spectra throughout adiabatic
clocking for clocking signals with different maxima.}
\par\end{center}{\small \par}

\noindent When $\gamma_{max}=0.1$ there is a level crossing between
the ground state (dashed) and first excited state (solid) which enhances
non-adiabatic transitions, so we adopt $\gamma_{max}=0.5$ as a practical
value. We can then solve the time dependent Schrodinger equation for
$|\Psi(t_{f})\rangle$ starting from the two different initial states
$|0\thinspace0\thinspace0\thinspace0\rangle$ and $|0\thinspace1\thinspace1\thinspace0\rangle$
and evaluate the transition probabilities to the various possible
wire final states. Starting from the state $|0\thinspace0\thinspace0\thinspace0\rangle$,
the switched wire ends up in its new ground state $|1\thinspace1\thinspace1\thinspace1\rangle$
with probability 0.9858. Furthermore, summing over transition probabilities,
the likelihood that the last cell in the wire is in the $|1\rangle$
state afterswitching is 0.9868. On the other hand, starting from the
$|0\thinspace1\thinspace1\thinspace0\rangle$ excited state, several
non-adiabatic transitions to other excited states occur. However,
summing over transition probabilities to states with last cell polarized
in the $|1\rangle$ state, we obtain 0.9868 again. This result is
summarized in Fig. 4.

\begin{center}
\includegraphics[scale=0.35]{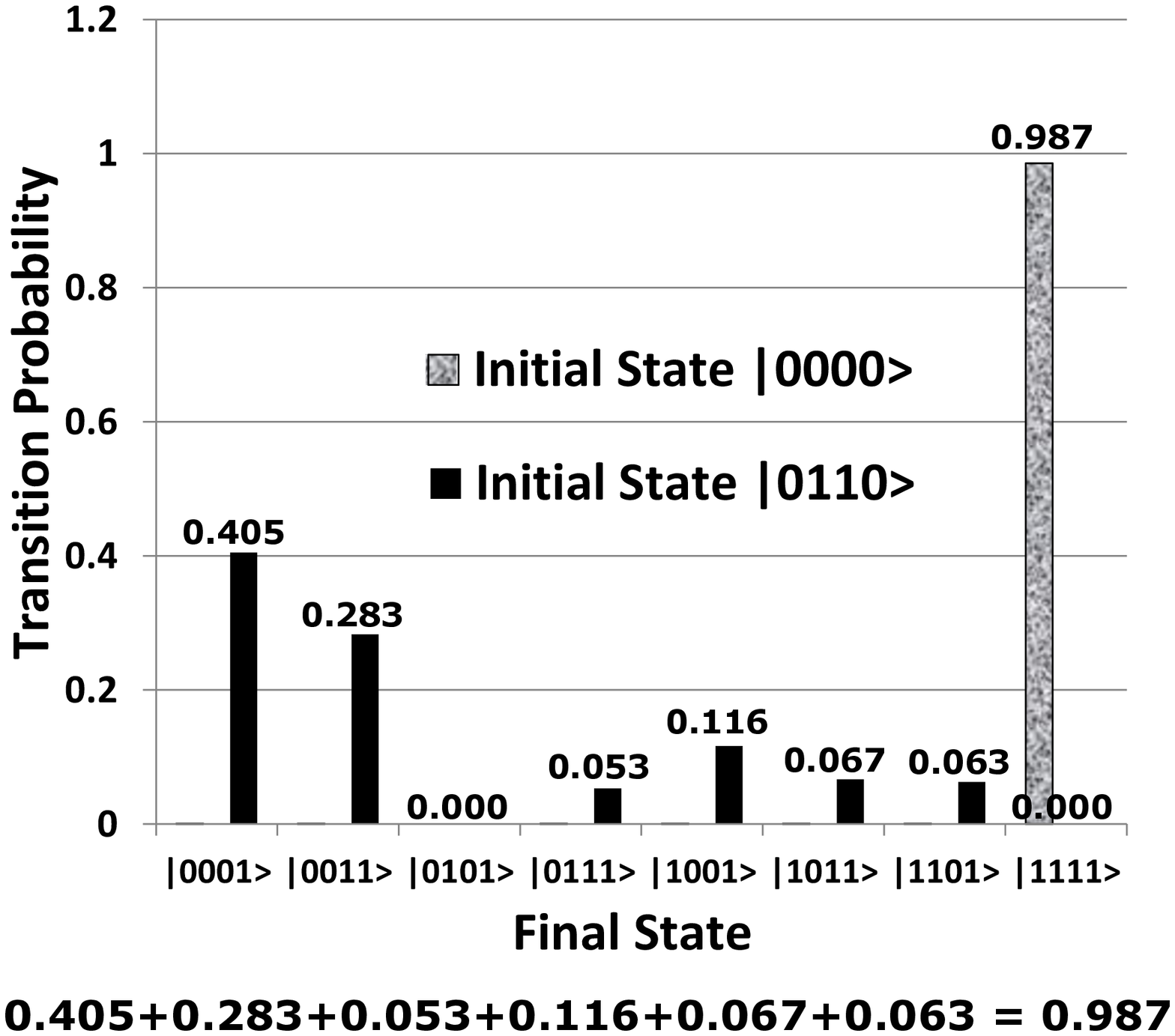}
\par\end{center}

\begin{center}
{\small{}Fig. 4 Transition probabilities to various wire states with
output polarization 1 after adiabatic switching is completed. The
two shades represent different initial states. In both cases the total
probability is the same.}
\par\end{center}{\small \par}

\noindent This result generalizes to wires of arbitrary length, kink
energies, choice of input polarization, and clocking functions. It
also holds for singly-branched inverters of the form shown in Fig.
5. In the case of adiabatic switching, the number of kinks in the
wire between initial and final wire states is conserved with high
probability as in Fig. 4. This is not a consequence of adiabaticity
alone, but is particular to the structure of the QCA wire Hamiltonian.

For more complex circuits such as the majority and doubly-branched
inverter gates shown in Fig. 6, the result does not hold. In this
figure, the ground and excited initial states of majority and inverter
gates are indicated together with the expected polarizations of the
output cell after the leftmost input cell is adiabatically switched
from $|0\rangle$ to $|1\rangle$. When a switching time of $t_{f}=30\hbar/E_{k}$
is used for the majority gate and a switching time of $t_{f}=60\hbar/E_{k}$
is used for the inverter, the respective expected values of the output
cell polarizations (ie. Pauli operator) after switching from the ground
states are 0.9852 and 0.8569. Both these polarizations correspond
to the desired $|1\rangle$ output state. However, flipping states
of internal cells in the initial configuration of either circuit to
generate excited states result in negative output polarizations after
switching, corresponding to the undesired $|0\rangle$ output state.
It follows that excited states of these circuits cannot be reliably
used for computation. 

\begin{center}
\includegraphics[scale=0.2]{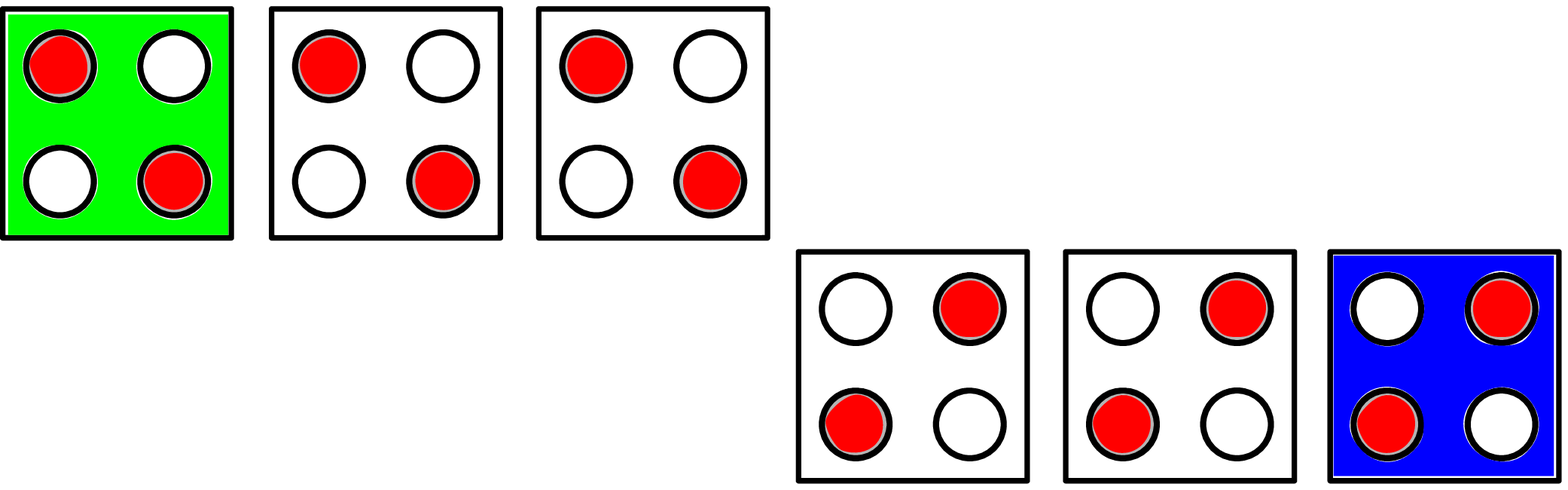}
\par\end{center}

\begin{center}
{\small{}Fig. 5 Singly-branched inverter circuit with similar QCA
Hamiltonian structure as a wire.}
\par\end{center}{\small \par}

\begin{center}
\includegraphics[scale=0.18]{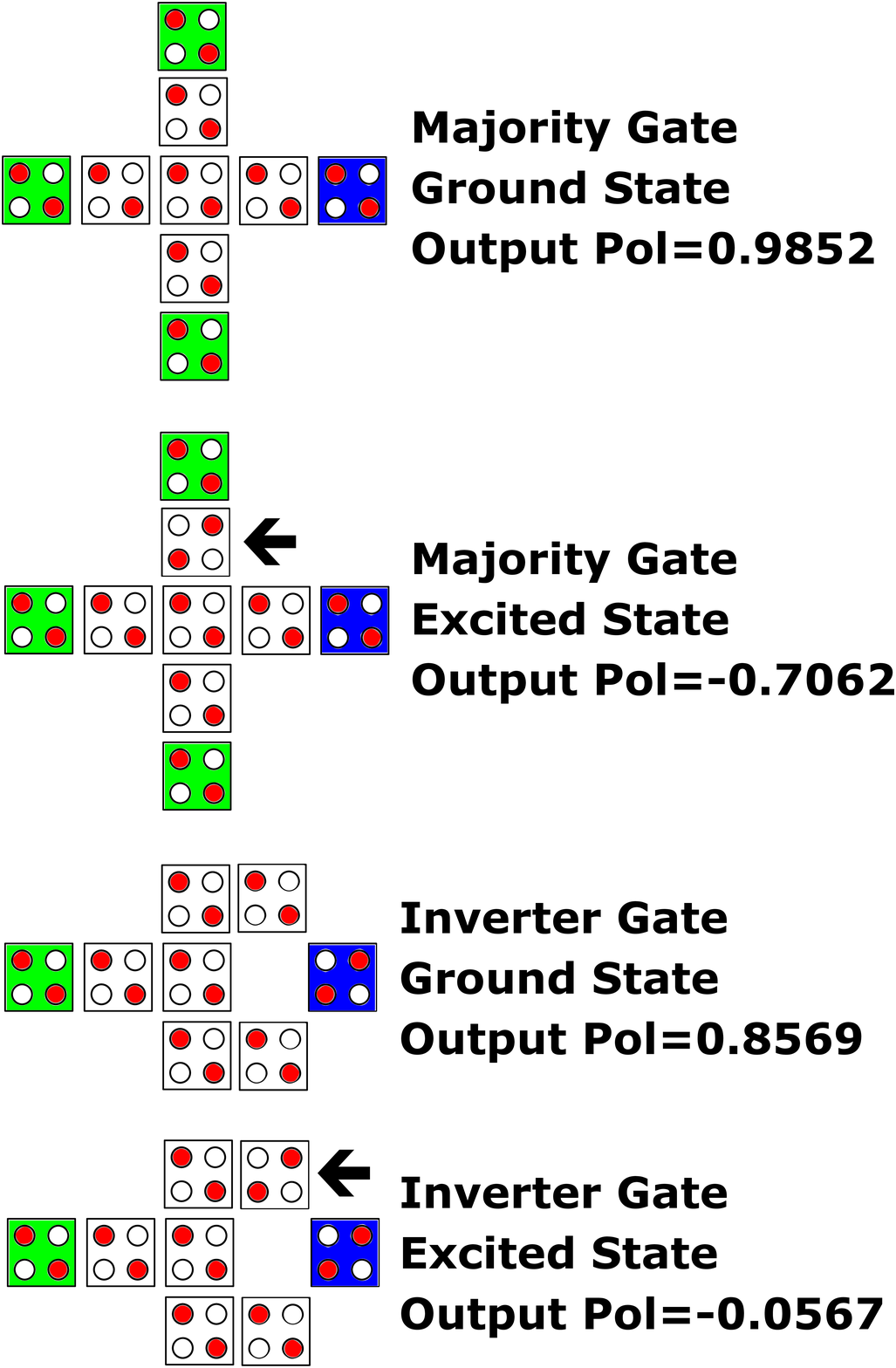}
\par\end{center}

\begin{center}
{\small{}Fig. 6 Schematics of QCA majority and inverter gate circuits.}
\par\end{center}{\small \par}

\section{Derivation}

In general, for an $N+1$ cell wire with a single input cell and general
clocking, we can write the wire Hamiltonian as:

\begin{eqnarray}
H_{wire}(t) & = & -\frac{E_{k}}{2}P_{in}(t)\sigma_{z}^{1}+\frac{E_{k}}{2}\sum_{i=1}^{N}\gamma^{i}(t)\sigma_{x}^{i}\\
 &  & -\frac{E_{k}}{2}\sum_{i=1}^{N-1}\sigma_{z}^{i}\sigma_{z}^{i+1}.\nonumber 
\end{eqnarray}

\noindent Information is transmitted along the wire by adjusting the
input polarization $P_{in}(t)$ and clocking signals $\gamma^{i}(t)$.
The input polarization affects the leftmost cell 1, and the wire output
is the $N^{th}$ cell. The coherent evolution of the wire is governed
by the time-dependent Schrodinger equation, or equivalently, the Von
Neumann equation {[}8{]}:

\begin{equation}
\dot{\rho}=-\frac{i}{\hbar}\left[H_{wire}(t),\rho\right],
\end{equation}

\noindent where $\rho$ is the density matrix of the wire describing
a pure state. The expected polarization of the wire output is given
by:

\begin{equation}
P_{out}(t)=Tr\left(\sigma_{z}^{N}\rho\right).
\end{equation}

\noindent This output polarization can be switched adiabatically by
slowly varying the input polarization and clocking signals. We now
prove that $P_{out}(t_{f})$ only depends on the initial expected
values of a particular set of operators, so that any state with the
same expectation values at $t=0$ will evolve into a state with the
same output polarization. 

To do this, for reasons that will become clear, rename:

\begin{equation}
Z_{0}(t)=P_{out}(t)=Tr\left(\sigma_{z}^{N}\rho\right),
\end{equation}

\noindent so that:

\begin{eqnarray}
\dot{Z}_{0}(t) & = & Tr\left(\sigma_{z}^{N}\dot{\rho}\right),\\
 & = & -\frac{i}{\hbar}Tr\left(\sigma_{z}^{N}\left[H_{wire}(t),\rho\right]\right),\nonumber \\
 & = & -\frac{i}{\hbar}Tr\left(\left[\sigma_{z}^{N},H_{wire}(t)\right]\rho\right),\nonumber \\
 & = & -\frac{i}{\hbar}Tr\left(\left(E_{k}\gamma^{N}(t)i\sigma_{y}^{N}\right)\rho\right),\nonumber \\
 & = & \frac{E_{k}}{\hbar}\gamma^{N}(t)Tr\left(\sigma_{y}^{N}\rho\right).\nonumber 
\end{eqnarray}

\noindent Similarly, letting $Y_{0}(t)=Tr\left(\sigma_{y}^{N}\rho\right)$:

\begin{eqnarray}
\dot{Y}_{0}(t) & = & Tr\left(\sigma_{y}^{N}\dot{\rho}\right),\\
 & = & -\frac{i}{\hbar}Tr\left(\left[\sigma_{y}^{N},H_{wire}(t)\right]\rho\right),\nonumber \\
 & = & -\frac{i}{\hbar}Tr\left(\left(-E_{k}\gamma^{N}(t)i\sigma_{z}^{N}-E_{k}\sigma_{z}^{N-1}i\sigma_{x}^{N}\right)\rho\right),\nonumber \\
 & = & -\frac{E_{k}}{\hbar}\gamma^{N}(t)Tr\left(\sigma_{z}^{N}\rho\right)-\frac{E_{k}}{\hbar}Tr\left(\sigma_{z}^{N-1}\sigma_{x}^{N}\rho\right),\nonumber 
\end{eqnarray}

\noindent and more generally, defining:

\begin{eqnarray}
Z_{i}(t) & = & Tr\left(\sigma_{z}^{N-i}\sigma_{x}^{N-i+1}...\sigma_{x}^{N}\rho\right),\\
Y_{i}(t) & = & Tr\left(\sigma_{y}^{N-i}\sigma_{x}^{N-i+1}...\sigma_{x}^{N}\rho\right),
\end{eqnarray}

\noindent we find:

\begin{eqnarray}
\dot{Z}_{i}(t) & = & \frac{E_{k}}{\hbar}\gamma^{N-i}(t)Y_{i}(t)+\frac{E_{k}}{\hbar}Y_{i-1}(t),\\
\dot{Y}_{i}(t) & = & -\frac{E_{k}}{\hbar}\gamma(t)Z_{i}(t)-\frac{E_{k}}{\hbar}Z_{i+1}(t).
\end{eqnarray}

\noindent These equations hold true except for $Y_{N-1}(t)$, where:

\begin{eqnarray}
\dot{Y}_{N-1}(t) & = & \frac{E_{k}}{\hbar}P_{in}(t)Z_{N}(t),\\
Z_{N}(t) & = & Tr(\sigma_{x}^{1}\sigma_{x}^{2}...\sigma_{x}^{N}\rho),
\end{eqnarray}

\noindent and:

\begin{equation}
\dot{Z}_{N}(t)=-\frac{E_{k}}{\hbar}P_{in}(t)Y_{N-1}(t).
\end{equation}

\noindent Since these equations form a system of linear ordinary differential
equations, it follows that all the expected values $\left\{ Z_{i}(t),Y_{i}(t)\right\} $
at any given time only depend on there values at $t=0$. In particular,
if we start out in a state where the output polarization is latched:
$\gamma^{N}(0)=0$, and we assume the polarization has a definite
value $P_{out}(0)=\pm1$, it follows $Z_{0}(0)=\pm1$, and all other
initial values are 0. All such states, regardless of whether or not
they are the ground state, evolve with the same output polarization.
Note that this derivation does not depend on the uniformity or sign
of the kink energy $E_{k}$ between nearest neighbors, and thus works
equally well for singly-branched inverters.

\section{Non-Nearest Neighbor Interactions}

Given that the derivation in the pevious section is particular to
the wire Hamiltonian, and this wire Hamiltonian is an approximation
that ignores effects such as non-nearest neighbor interactions, it
is important to consider the effects of these interactions on information
transmission. Therefore, a modification of the wire Hamiltonian taking
into consideration next-to-nearest neighbor interactions is made as
shown below for the four cell wire:

\begin{eqnarray}
H_{total}(t) & = & H_{wire}(t)+H_{I}(t),\\
H_{wire}(t) & = & -\frac{E_{k}}{2}P_{in}(t)\sigma_{z}^{1}+\frac{E_{k}}{2}\gamma(t)(\sigma_{x}^{i}+\sigma_{x}^{i}+\sigma_{x}^{i}+\sigma_{x}^{i})\nonumber \\
 &  & -\frac{E_{k}}{2}(\sigma_{z}^{1}\sigma_{z}^{2}+\sigma_{z}^{2}\sigma_{z}^{3}+\sigma_{z}^{3}\sigma_{z}^{4}),\\
H_{I}(t) & = & -\frac{E_{k}}{64}(\sigma_{z}^{1}\sigma_{z}^{3}+\sigma_{z}^{2}\sigma_{z}^{4}).
\end{eqnarray}

\noindent In this Hamiltonian, the kink energy between next-to-nearest
neighbor cells is lower by a factor of $2^{-5}=1/32$ over the neighboring
interaction due to the quadrupole-quadrupole interaction between QCA
cells. Using the corresponding Hamiltonian for 6 and 7 cell wires,
the following table shows the computed output cell polarizations for
$60\hbar/E_{k}$ adiabatic switching cycles starting from different
initial wire states next to their corresponding unperturbed values:

\begin{center}
{\small{}TABLE I: Expected output cell polarization after adiabatic
switching from different initial wire states. }
\par\end{center}{\small \par}

\begin{center}
\begin{tabular}{|c|c|c|}
\hline 
Initial Wire State & $H_{I}(t)=0$ & $H_{I}(t)\neq0$\tabularnewline
\hline 
\hline 
{[}000000{]} - 6 cell & 0.9932 & \textbf{0.9572}\tabularnewline
\hline 
{[}000010{]} - 6 cell & 0.9932 & 0.9798\tabularnewline
\hline 
{[}000100{]} - 6 cell & 0.9932 & 0.9854\tabularnewline
\hline 
{[}010000{]} - 6 cell & 0.9932 & 0.9794\tabularnewline
\hline 
{[}010100{]} - 6 cell & 0.9932 & 0.9949\tabularnewline
\hline 
{[}101010{]} - 6 cell & 0.9932 & 0.9966\tabularnewline
\hline 
{[}0000000{]} - 7 cell & 0.9784 & \textbf{0.8795}\tabularnewline
\hline 
{[}0011000{]} - 7 cell & 0.9784 & 0.9492\tabularnewline
\hline 
{[}1001000{]} - 7 cell & 0.9784 & 0.9797\tabularnewline
\hline 
{[}1101100{]} - 7 cell & 0.9784 & 0.9753\tabularnewline
\hline 
\end{tabular}
\par\end{center}

\noindent This data shows that with non-nearest neighbor interactions,
the ground state and excited state final cell polarizations after
switching are no longer equivalent, but are small perturbations of
the identical polarization obtained in the absence of such interactions.
Furthermore, in all cases simulated, including all excited states
of a 6 cell wire, the excited state performs better than the ground
state. Fig. 7 shows a plot of the difference between the cell 6 output
polarizations with next-to-nearest neighbor interaction and without
vs time (units $\hbar/E_{k}$) for the ground state and two excited
states:

\begin{center}
\includegraphics[scale=0.55]{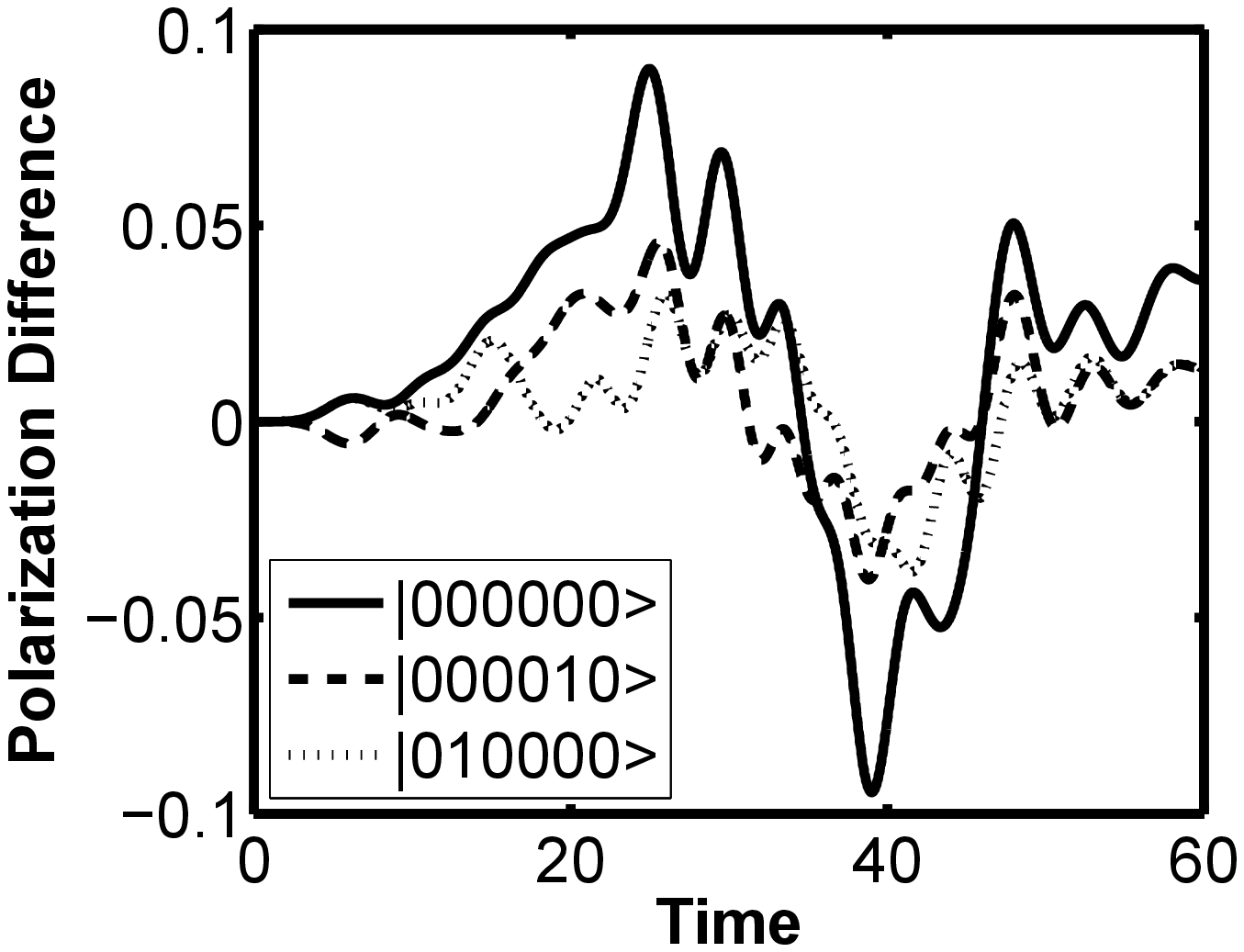}
\par\end{center}

\begin{center}
{\small{}Fig. 7 Diference between cell 6 output polarizations of wires
with and without next-to-nearest neighbors for starting from the ground
state and two excited states.}
\par\end{center}{\small \par}

\noindent In this figure, the fact that the ground state polarization
difference at the $t=60\hbar/E_{k}$ end of switching is greater than
the excited state polarization difference accounts for the better
performance of the excited states. It also appears that the excited
state polarization differences roughly mimic the ground state polarization
differences but are scaled down.

To get a sense of why this is true, we can write the density matrix
$\rho$ which evolves under the action of the Hamiltonian $H_{total}$
as the sum of two pieces:

\begin{equation}
\rho(t)=\rho_{wire}(t)+\rho_{I}(t),
\end{equation}

\noindent where:

\begin{eqnarray}
\dot{\rho}_{wire}(t) & = & -\frac{i}{\hbar}\left[H_{wire}(t),\rho_{wire}(t)\right],\\
\rho_{wire}(0) & = & \rho(0).
\end{eqnarray}

\noindent From this it follows from the Von Neumann equation for $\rho$
that:

\begin{eqnarray}
\dot{\rho}_{wire}+\dot{\rho}_{I} & = & -\frac{i}{\hbar}\left[H_{wire}+H_{I},\rho_{wire}+\rho_{I}\right],\\
\dot{\rho}_{I} & = & -\frac{i}{\hbar}\left[H_{wire}+H_{I},\rho_{I}\right]-\frac{i}{\hbar}\left[H_{I},\rho_{wire}\right],\\
\rho_{I}(0) & = & 0,
\end{eqnarray}

\noindent and by similar manipulations to those of the previous section:

\begin{eqnarray}
\frac{d}{dt}Tr(\sigma_{z}^{N}\rho_{I}) & = & \frac{E_{k}}{\hbar}Tr(\sigma_{y}^{N}\rho_{I})\gamma^{N}(t),\\
Tr(\sigma_{z}^{N}\rho_{I}(0)) & = & 0.
\end{eqnarray}

\noindent This equation is of interest because $Tr(\sigma_{z}^{N}\rho_{I})$
is the change to the expected polarization of the $N^{th}$ cell in
the wire induced by non-nearest neighbor interactions. Its initial
value is zero, and its rate of change is proportional to the y-polarization
of the cell. If we assume that $\rho_{I}$ is negligible compared
to $\rho_{wire}$, we can further approximate:

\begin{equation}
\frac{d}{dt}Tr(\sigma_{y}^{N}\rho_{I})\approx-\frac{E_{k}}{\hbar}Tr(\sigma_{z}^{N-2}\sigma_{x}^{N}\rho_{wire}(t)),
\end{equation}

\noindent and the function of time on the right depends on the initial
state of the wire. It appears this function is highly oscillatory
for excited states while not so for the ground state, so $Tr(\sigma_{y}^{N}\rho_{I})$
is correspondingly smaller for excited states, as is the final output
polarization $Tr(\sigma_{z}^{N}\rho_{I})$. Of course, this is only
heuristic argument, and a more rigorous theoretical derivation remains
outstanding.

\section{Conclusions}

Simulation of QCA wires suggests that initial occupation of the wire
ground state is not necessary for reliable information transmission,
despite the occurrence of non-adiabatic transitions in the case where
the initial state is excited. In the case where non-nearest neighbor
interactions are neglected from the wire Hamiltonian, it is derived
that there is identical behavior of information transmission along
QCA wires and singly-branched inverters occupying their ground or
excited states. On the other hand, other simple circuits such as majority
and double-branched inverter gates do not share this property. Therefore,
in a QCA circuit that has been divided into separate individually
clocked zones, this suggests that wires used to relay information
around the circuit are still functional after environmental excitation
of their quantum state, while gates are not. This information is of
relevance in designing QCA circuits with appropriate error correction,
as it suggests gates performing computations should be contained in
smaller clocking zones than wires relaying information for optimized
circuit performance.

\section*{Acknowlegment}

\noindent This work was supported by an NSERC Engage grant in parternship
with Quantum Silicon Inc.

\section{References}

\noindent {[}1{]} C. Lent, P. Tougdaw, ``A device architecture for
computing with quantum dots'', \textit{Proc. IEEE,} vol. 85, no.
4, pp. 541 - 557, 1997.

\noindent {[}2{]} J. Timler, C. Lent, ``Power gain and dissipation
in quantum-dot cellular automata'', \textit{Journal of Applied Physics},
vol. 91, no. 2, pp. 823 - 831, 2002.

\noindent {[}3{]} T. Kato, \textquotedbl{}On the adiabatic theorem
of quantum mechanics\textquotedbl{}. \textit{Journal of the Physical
Society of Japan,} vol. 5, no. 6, pp. 435 - 439, 1950.

\noindent {[}4{]} L. Livadaru, P. Xue, Z. Shaterzadeh-Yazdi, G. DiLabio,
J. Mutus, J. Pitters, B. Sanders, R. Wolkow, ``Dangling-bond charge
qubit on a silicon surface'', \textit{New Journal of Physics}, vol.
12, no. 8, 083018, 2010.

\noindent {[}5{]} S. Srivastava, S. Bhanja, \textquotedbl{}Hierarchical
probabilistic macromodeling for QCA circuits.\textquotedbl{} \textit{IEEE
Transactions on Computers,} vol. 56, no. 2, pp. 174 - 190, 2007.

\noindent {[}6{]} C. Zener, ``Non-adiabatic crossing of energy levels'',
\textit{Proceedings of the Royal Society of London. Series A, Containing
Papers of a Mathematical and Physical Character, }vol. 137, no. 833,
pp. 696 - 70, 1932.

\noindent {[}7{]} F. Karim, A. Navabi, K. Walus, A. Ivanov \textquotedbl{}Quantum
mechanical simulation of QCA with a reduced Hamiltonian model\textquotedbl{},
\textit{Proceedings of the 8th IEEE Conference on Nanotechnology,
2008}. 

\noindent {[}8{]} G. Mahler, and A. Volker, \textit{\small{}Quantum
networks, Dynamics of open nanostructures}, Springer-Verlag Berlin
Heidelberg (1998).
\end{document}